\documentclass[aps,prd,nofootinbib,superscriptaddress,twocolumn]{revtex4}

\usepackage{amsmath,amsfonts,amssymb}
\usepackage{booktabs}
\usepackage{enumitem}
\usepackage{graphicx}
\usepackage{siunitx}
\usepackage{xcolor}
\usepackage{verbatim}
\usepackage[colorlinks=true,urlcolor=black,citecolor=blue,linkcolor=black]{hyperref}
\usepackage{enumitem}
\setlist[itemize]{leftmargin=*}

\newcommand{\ie}{{\it i.e.}}

\newcommand{\be}{\begin{equation}}
\newcommand{\ee}{\end{equation}}
\newcommand{\br}{\begin{eqnarray}}
\newcommand{\bea}{\begin{eqnarray}}
\newcommand{\eea}{\end{eqnarray}}
\newcommand{\er}{\end{eqnarray}}
\newcommand{\ba}{\begin{array}}
\newcommand{\ea}{\end{array}}
\newcommand{\bi}{\begin{itemize}}
\newcommand{\ei}{\end{itemize}}
\newcommand{\bn}{\begin{enumerate}}
\newcommand{\en}{\end{enumerate}}
\newcommand{\bc}{\begin{center}}
\newcommand{\ec}{\end{center}}

\newcommand{\gsim}{\lower.7ex\hbox{$\;\stackrel{\textstyle>}{\sim}\;$}}
\newcommand{\lsim}{\lower.7ex\hbox{$\;\stackrel{\textstyle<}{\sim}\;$}}

\begin{document}


\author{Kristjan Kannike}
 \email{kristjan.kannike@cern.ch}
\author{Niko Koivunen}
\email{niko.koivunen@kbfi.ee}
\author{Martti Raidal}
 \email{martti.raidal@cern.ch}
\affiliation{NICPB, R\"{a}vala 10, Tallinn 10143, Estonia}

\title{Principle of Multiple Point Criticality in Multi-Scalar Dark Matter Models}

\begin{abstract}
The principle of multiple point criticality (PMPC), which allowed the prediction of the Higgs boson mass before its discovery, has so far been applied to
 radiatively generated vacua. If this principle is fundamental, following from some presently unknown underlying physics, the PMPC must
apply to all vacua, including the multiple vacua of multi-scalar models dominated by {\it tree-level} terms. We first motivate this idea and then exemplify it by applying the 
PMPC to various realizations of singlet scalar dark matter models. We derive constraints on the dark matter properties from the requirement of degenerate vacua
and show that some scalar dark matter models are ruled out by the PMPC, while in others the allowed parameters space is constrained.
\end{abstract}

\maketitle

\section{Introduction}

The discovery of the standard model (SM) Higgs boson~\cite{Chatrchyan:2012xdj,Aad:2012gk}, 
the only known elementary scalar particle, has been the main result of the Large Hadron Collider
physics program. Remarkably, the measured Higgs boson mass was actually predicted with a good accuracy already in 1995~\cite{Froggatt:1995rt}  by
applying the principle of multiple point criticality (PMPC)~\cite{Bennett:1993pj,Bennett:1996vy,Bennett:1996hx} to the Higgs boson effective potential. 
Today the SM Higgs criticality has been studied to three-loop accuracy, indicating that the present data prefers a meta-stable Higgs vacuum 
sufficiently close to criticality~\cite{Buttazzo:2013uya}.\footnote{This statement depends most strongly on the precise value of the top quark Yukawa coupling, as the critical point remains within $3\sigma$ uncertainty of $y_t$. Since $y_t$ is not a directly measured quantity, the systematic uncertainty related to its derivation from collider top mass measurements may be presently underestimated~\cite{Corcella:2019tgt}.}
This fact is the most interesting unexplained experimental outcome of collider physics so far.
The PMPC has also been applied to explain dark matter with only the SM particle content \cite{Froggatt:2005js, Froggatt:2014tua, Froggatt:2005fk}. 

Within the SM, there is no explanation as to why the SM parameters -- the gauge, Yukawa and scalar quartic couplings -- must take values leading to the Higgs criticality.
Therefore, in the original proposal~\cite{Froggatt:1995rt}, the existence of multiple degenerate vacua was argued to be analogous to the 
simultaneous existence of multiple phases in thermodynamics. This statement is purely empirical  and cannot follow solely from the SM physics, nevertheless
it may have some support from statistical systems in solid state physics~\cite{Volovik:2003hn}.
To derive the PMPC from fundamental physical principles, Higgs criticality has been attributed to the proposed asymptotic safety of quantum gravity~\cite{Shaposhnikov:2009pv}.
More recently Dvali~\cite{Dvali:2011wk} revived the old argument by Zeldovich~\cite{Zeldovich:1974py,Kobzarev:1974cp}  that physical systems with multiple vacua must all be degenerate because of the consistency of quantum field theory -- tunneling from a Minkowski vacuum to a lower one must be exponentially fast as in any theory possessing ghosts. 
The Zeldovich-Dvali argument contradicts Coleman's semi-classical computation in the Euclidian space that the formation of vacuum bubbles is exponentially slow~\cite{Coleman:1977py},  and has also been challenged by studies of different physical systems~\cite{Garriga:2011we}. However, attempts to generalize Coleman's computation to the realistic case with zero-mass vacuum bubbles, which is the origin of the divergence, have failed and the Zeldovich-Dvali argument has not been disproved~\cite{Gross:2020tph}. Altogether, the underlying fundamental physics behind the PMPC remains presently unclear: we accept it on empirical grounds.

Nevertheless, the phenomenological success of PMPC has motivated different studies of scalar potentials in extended models. Already the very first papers
after the Higgs boson discovery argued that if the SM is extended by a singlet scalar, the Higgs vacuum stability as well as criticality can be achieved due to the radiative corrections of the singlet~\cite{Kadastik:2011aa}. Later the PMPC has been applied to the singlet~\cite{Haba:2014sia,Kawana:2014zxa,Haba:2016gqx,McDowall:2019knq,Hamada:2020wjh} and to the two Higgs doublet (2HDM)~\cite{Froggatt:2004st,Froggatt:2006zc,Froggatt:2008am,McDowall:2018ulq,Maniatis:2020wfz} extensions of the SM.
The Higgs criticality may play an important role in understanding Higgs inflation~\cite{Bezrukov:2007ep}. We do not make an attempt to form a comprehensive list of all possible applications of PMPC and limit ourselves to the above examples.

The aim of this work is to point out that so far all the applications of PMPC to scalar potentials, including the examples given above, deal with radiatively generated
multiple minima. However, if the degeneracy of multiple vacua is a fundamental property of Nature, all vacua of multi-scalar models -- including the ones 
dominated by {\it tree-level} potential terms -- should be degenerate. We first argue that this formulation of PMPC should be correct, and after that exemplify its physics potential, analyzing the vacuum solutions of the scalar singlet dark matter (DM) model~\cite{Silveira:1985rk,McDonald:1993ex,Burgess:2000yq,Cline:2013gha} (see~\cite{Arcadi:2019lka} for a review).
We present three simple scalar DM models utilizing the PMPC: I) $\mathbb{Z}_2$-symmetric real scalar DM model, II) $\mathbb{Z}_2$-symmetric pseudo-Goldstone DM (pGDM) model and III) $\mathbb{Z}_3$-symmetric pseudo-Goldstone DM model.\footnote{In a more complicated pGDM model, all these global symmetries can be broken, save for the CP-like $S \to S^{*}$ that stabilizes DM \cite{Alanne:2020jwx}.} Our results will show that the allowed parameter spaces of those models are constrained and the PMPC 
will, in principle, be testable if the DM of Universe is the singlet scalar. Generalizing this result, our work demonstrates that the PMPC might be a useful tool to discriminate
between different models of new physics beyond the SM.

This work is organized as follows. In Section~\ref{sec:PMPC} we generalize the concept of PMPC to the case of multiple scalar fields. In Sections~\ref{sec:model1},~\ref{sec:model2} and~\ref{sec:model3} we analyze our example models. We conclude and discuss perspectives in Section~\ref{sec:conc}.

\section{Tree-level PMPC}
\label{sec:PMPC}

So far the PMPC has been used to constrain model parameters by requiring that a radiatively generated minimum of the
scalar potential is degenerate with the electroweak vacuum. This was the case with the SM as well as with the scalar singlet and doublet
extensions of the SM. Clearly, in multi-scalar models there can be several minima of the 
scalar potential already at tree level. If the PMPC follows from some underlying fundamental physics principle, all those vacua must be degenerate as well.
This is the concept we adopt and study in this work. Although we do not know which physics principle is behind the PMPC, we exemplify with the following discussion
that such principles may exist.

As Zeldovich pointed out in early 70ies~\cite{Zeldovich:1974py,Kobzarev:1974cp}, if there exist two minima of the scalar potential with $V=0$ and $V<0$,
the vacuum decay bubble with mass $m = 0$ can appear with any initial velocity, giving rise to a divergent boost integral. Dvali, who apparently was not aware of the Zeldovich paper, independently make this claim in Ref.~\cite{Dvali:2011wk}. Notice that this problem has a close analogue in the theories with ghosts, \ie, to particles with negative energy. These theories are widely believed to be pathological precisely because of the exponentially fast vacuum decay, analogously to the Lorentz-invariant decay of the Minkowski vacuum conjectured by Zeldovich and Dvali, rendering this configuration unphysical. Indeed, in theories of ghosts, the same divergence was found recently~\cite{Gross:2020tph}. There is no such a problem with the pair creation of ordinary particles.

This argument leads to the PMPC in the region of the parameter space where there is a second minimum that can be lower than our Minkowski-like vacuum. The decay of the present minimum to the lower one would have an infinite rate and therefore be unphysical. This decay can be prevented in presence of gravity, which will make the required radius of the critical bubble larger by the Coleman-de Luccia suppression \cite{Coleman:1980aw}. If the energy difference between the two minima is small enough, the vacuum decay does not happen. In the limit of weak gravity the requirement of no tunneling means that the two minima must be nearly degenerate as required by the PMPC. This does not exclude, though, the existence of minima higher than ours.

Needless to say, this claim contradicts Coleman's result that the vacuum decay rate is finite and exponentially suppressed~\cite{Coleman:1977py}. The Zeldovich-Dvali claim was also criticized in Ref.~\cite{Garriga:2011we} which,
however, studies a completely different physical situation -- particle pair production in uniform electric field. Whether this set-up is analogous to the case of 
vacuum decay is far from being obvious. It seems that the main source of different opinions is whether the divergent boost integral has to be taken or not,
and there is no definitive answer at the moment. 

Nevertheless, this discussion demonstrates that there exist attempts to derive PMPC from general physics principles.
The PMPC in this, broader formulation has been used to criticize certain supersymmetry breaking scenarios~\cite{Bajc:2011iu}. In addition, the fact that the SM Higgs vacuum is metastable, that is very close to criticality~\cite{Buttazzo:2013uya} may not be a coincidence, but can be a hint for the existence of the PMPC with a degenerate minimum near the Planck scale as in the original proposal \cite{Froggatt:1995rt}.

In this work we perform phenomenological analyses of implications of tree-level PMPC on multi-scalar DM models.

\section{Model I: $\mathbb{Z}_2$ real singlet}
\label{sec:model1}

The real scalar singlet DM model extends the SM with one real scalar singlet which is stabilized by a $\mathbb{Z}_2$ symmetry. 
This is among the most popular DM models whose phenomenology has been thoroughly studied~\cite{Davoudiasl:2004be,Ham:2004cf,OConnell:2006rsp,Patt:2006fw,Profumo:2007wc,Barger:2007im,He:2007tt,He:2008qm,Yaguna:2008hd,Lerner:2009xg,Farina:2009ez,Goudelis:2009zz,Profumo:2010kp,Guo:2010hq,Barger:2010mc,Arina:2010rb,Bandyopadhyay:2010cc,Drozd:2011aa,Djouadi:2011aa,Low:2011kp,Mambrini:2011ik,Espinosa:2011ax,Mambrini:2012ue,Djouadi:2012zc,Cheung:2012xb,Cline:2013gha,Urbano:2014hda,Endo:2014cca,Feng:2014vea,Duerr:2015bea,Duerr:2015mva,Beniwal:2015sdl,Cuoco:2016jqt,Escudero:2016gzx,Han:2016gyy,He:2016mls,Ko:2016xwd,Athron:2017kgt,Ghorbani:2018yfr}. Nevertheless, we shall show that applying the PMPC to this model scalar potential will open new perspectives in
understanding the allowed parameter space of the model.

The scalar potential of the model is given by
\be
V=\mu_h^2 \lvert H \rvert^{2} + \frac{\mu_s^2}{2}s^2+\lambda_h \lvert H \rvert^{4}+\frac{\lambda_s}{4}s^4+\frac{\lambda_{hs}}{2} \lvert H \rvert^{2} s^2,
\ee
where $H$ is the SM Higgs doublet in unitary gauge:
\be
H=\frac{1}{\sqrt{2}}\left(\begin{array}{c}
0\\
h+v_h
\end{array}
\right).
\ee
The Higgs vacuum expectation value is $v_h=246$ GeV.
The scalar $s$, as the  dark matter candidate, must not acquire a non-zero VEV: then the scalar $s$ does not mix with the Higgs and remains stable due to the $\mathbb{Z}_2$ symmetry. The freeze-out of the $s$ is independent of the self-coupling of the scalar singlet $\lambda_s$ and its relic abundance is therefore set only by the Higgs portal coupling $\lambda_{hs}$. 

The potential may have three types of minima in the field space $(h,s)$: the Higgs vacuum $(v_h,0)$, singlet vacuum $(0,v_s)$ and a general or mixed vacuum $(v'_h,v'_s)$. Presently, the Universe is in the Higgs vacuum. According to the PMPC these minima are to be degenerate, if they exist simultaneously. Let us examine the three vacuum configurations more closely. 

\subsection{The Higgs vacuum configuration $(v_h,0)$}
The Higgs vacuum $(v_h,0)$, in which we live, satisfies the following minimization condition:
\be
\mu_h^2=-\lambda_h v_h^2.\label{Our vacuum minimization condition}
\ee 
The scalar mass matrix is given by
\be
\mathcal{M}^2=\left(\begin{array}{cc}
2\lambda_h v_h^2 & 0\\
0 & \frac{1}{2}\lambda_{hs} v_h^2+\mu_s^2
\end{array}
\right).
\ee

\subsection{The singlet vacuum configuration $(0,v_s)$}
The minimization condition for the singlet vacuum $(0,v_s)$ is given by
\be
\mu_s^2=-\lambda_s v_s^2.\label{singlet vacuum minimization condition}
\ee

In this case, the scalar mass matrix is given by
\be
\mathcal{M}^2=\left(\begin{array}{cc}
\frac{1}{2}\lambda_{hs} v_s^2+\mu_h^2  & 0\\
0 & 2\lambda_s v_s^2
\end{array}
\right).
\ee

\subsection{The mixed vacuum configuration $(v'_h,v'_s)$}
The minimization conditions for the mixed vacuum $(v'_h,v'_s)$  are given by
\bea
\mu_h^2 & = & -\lambda_h {v'_h}^2-\frac{1}{2}\lambda_{hs}{v'_s}^2\\
\mu_s^2 & = & -\lambda_{s} {v'_s}^2-\frac{1}{2}\lambda_{hs} {v'_h}^2
\eea
The scalar mass matrix is given by
\be
\mathcal{M}^2=\left(\begin{array}{cc}
2\lambda_{h} {v'_h}^2  & \lambda_{hs} {v'_h}v'_s\\
\lambda_{hs} {v'_h}v'_s & 2\lambda_{s} {v'_s}^2
\end{array}
\right),
\ee
leading to mixing of the Higgs doublet with the real singlet.

\subsection{Comparison of minima}
The current minimum configuration is the Higgs vacuum $(v_h,0)$, which allows dark matter and correctly breaks the electroweak symmetry. If the other vacuum configurations are to exist simultaneously with the $(v_h,0)$, they must be degenerate. Imposing that the $(0,v_s)$ vacuum configuration is degenerate to $(v_h,0)$,
\be
V(h=v_h,s=0)= V(h=0,s=v_s),
\ee
leads to the condition
\be\label{the constraint for our vacuum VS singlet vacuum}
m_s^4-\lambda_{hs} m_s^2 v_h^2 -\lambda_s\lambda_h v_h^4+\frac{1}{4}\lambda_{hs}^2 v_h^4=0.
\ee

Demanding that the $(v'_h,v'_s)$ vacuum configuration be degenerate to the present Higgs vacuum $(v_h,0)$ leads to a contradiction.  Therefore, according to the PMPC, there can at most only be two degenerate vacua: the Higgs vacuum $(v_h,0)$ and the singlet vacuum $(0,v_s)$. 

The collider experiments impose constraints on the singlet $s$  when it is lighter than $m_h/2$. Then the Higgs decay width to the $ss$ final state is \cite{Cline:2013gha}
\be
\Gamma (h\to ss) =\frac{\sqrt{m_h^2-4m_s^2}}{32\pi m_h^2}\lambda_{hs}^2 v_h^2. 
\ee
In addition, the Higgs invisible branching ratio is then given by
\begin{equation}
\text{BR}_\text{inv} = \frac{\Gamma_{h \to ss}}{\Gamma_{h \to \text{SM}} + \Gamma_{h \to ss}},
\end{equation}
which is constrained to be below $0.24$ at $95\%$ confidence level \cite{Khachatryan:2016whc,ATLAS-CONF-2018-031} by direct measurements and below about $0.17$ by statistical fits of Higgs couplings \cite{Giardino:2013bma,Belanger:2013xza}, which excludes singlet masses below $53$ GeV.

\begin{figure*}[tb]
	\begin{center}
	\includegraphics[width=0.45\linewidth]{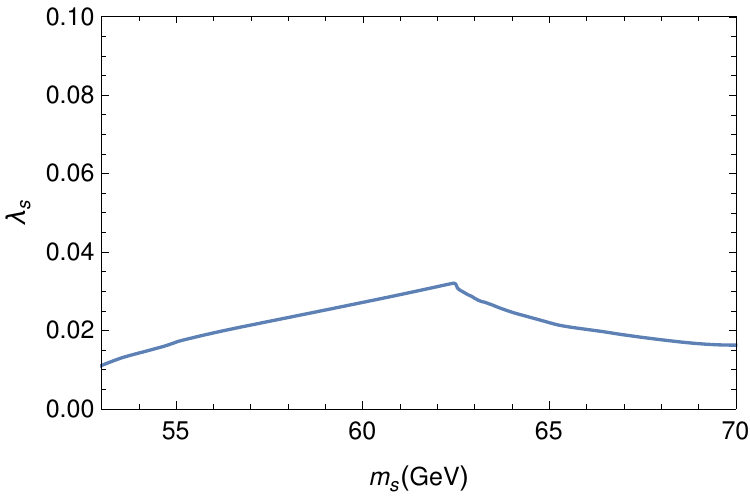}
	\includegraphics[width=0.45\linewidth]{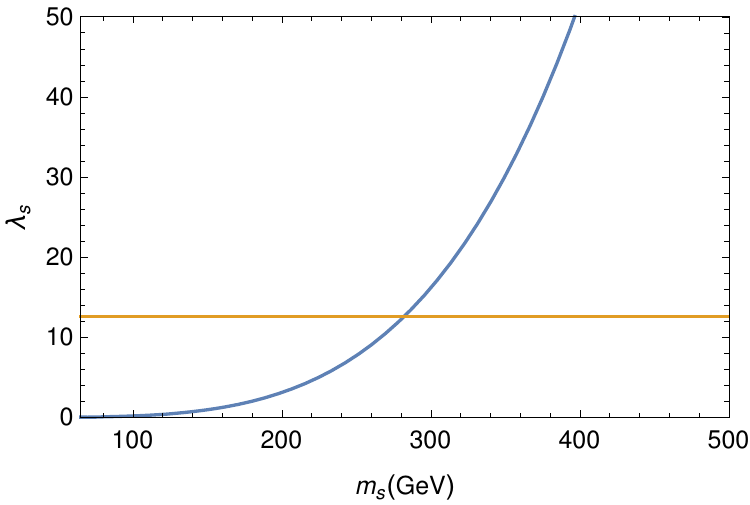}
	\end{center}
	\caption{\label{dark matter self-coupling as a function of dark matter mass} The value of $\mathbb{Z}_2$ real singlet DM self-coupling $\lambda_s$ as a function of DM mass according to PMPC in Eq. \eqref{the constraint for our vacuum VS singlet vacuum}. The portal coupling is fixed so that the correct relic abundance is obtained. Left panel: the self-coupling $\lambda_s$ near the Higgs resonance, $m_s\in [53,70]$ GeV. Right panel: the  self-coupling $\lambda_s$ away from the resonance, $m_s\in [65,1000]$ GeV.  }
\end{figure*}

The portal coupling $\lambda_{hs}$ is determined by demanding that the correct relic abundance $\Omega_{c} h^{2} = 0.120$ \cite{Aghanim:2018eyx} is produced in the freeze-out of the singlet. The PMPC in Eq.~(\ref{the constraint for our vacuum VS singlet vacuum}) allows us to determine the singlet self-coupling $\lambda_s$ as a function of the dark matter mass, as is presented in the Fig. \ref{dark matter self-coupling as a function of dark matter mass}. The self-coupling $\lambda_s$ becomes non-perturbative as the dark matter mass reaches electroweak scale. 

Direct detection also limits 
the dark matter mass so that the only allowed mass region is near the Higgs resonance. The PMPC thus predicts that the dark matter mass must be in the resonance region. 

\section{Model II: $\mathbb{Z}_2$ pseudo-Goldstone DM}  
\label{sec:model2}

This  model is an extension of SM with a  complex scalar $S$ carrying a global $U(1)$ charge.   The most general scalar potential invariant under global $U(1)$ transformation is given by
\be\label{}
\begin{split}
  V_0 &= \mu_h^2 \lvert H \rvert^{2} + \mu_s^2 \lvert S \rvert^{2} +\lambda_h \lvert H \rvert^{4}
  +\lambda_s \lvert S \rvert^{4} \\
  &+\lambda_{hs} \lvert H \rvert^{2} \lvert S \rvert^{2},
\end{split}
\ee
where $H$ is the SM Higgs doublet. 
The global $U(1)$ symmetry is spontaneously broken as the scalar $S$ acquires a non-zero vacuum expectation value. This would result in a massless Goldstone boson in the physical spectrum. In order to avoid this, the global $U(1)$ symmetry is explicitly broken into its discrete subgroup $\mathbb{Z}_2$ by the following soft term:
\be
V_{\mathbb{Z}_2}=-\frac{\mu^{\prime 2}}{4}S^2+\text{h.c},
\ee
so the full potential becomes
\be\label{Z2 scalar potential}
V=V_0+V_{\mathbb{Z}_2}.
\ee

The model was first considered in \cite{Barger:2008jx,Chiang:2017nmu}, but only in the Higgs resonance region. The suppression of the direct detection cross section of the pseudo-Goldstone DM was noted in \cite{Gross:2017dan}; there has been large interest in the phenomenology of the model \cite{Huitu:2018gbc,Azevedo:2018oxv,Azevedo:2018exj,Alanne:2018zjm,Cline:2019okt,Kannike:2019wsn}. The minimum structure and phase transitions of models with two singlets have been considered e.g. in \cite{Barger:2008jx,Chiang:2017nmu,Vieu:2018nfq,Ghorbani:2019itr}.

The parameter $\mu^{\prime 2}$ can be made real and positive by absorbing its phase into the field $S$. When $\mu^{\prime 2}$ is positive, then the VEV of $S$ is real. The reality of the VEV can be easily seen by writing the singlet in polar coordinates:
\be
S=\frac{1}{\sqrt{2}}r e^{i\theta}.
\ee
The soft term now becomes
\be
V_{\mathbb{Z}_2}=-\frac{\mu^{\prime 2}}{4}r^2\cos (2\theta).
\ee
The soft term is the only part of the potential that depends on the phase of the singlet $S$. Therefore the global minimum of the potential is the minimum of the soft term.  Because  $\mu^{\prime 2}>0$, the minimum is obtained when $\cos(2\theta)=1$. It follows that
\be
2\theta =\pm 2\pi n\Rightarrow \theta=0,\pi,
\ee
and therefore the VEV is real.

We parametrize the fields in the unitary gauge as:
\be
H=\frac{1}{\sqrt{2}}\left(\begin{array}{c}
0\\
h+v_h
\end{array}
\right),
\ee
and
\be
S=\frac{1}{\sqrt{2}}(v_s+s+i\chi).
\ee
The Higgs vacuum expectation value is $v_h=246$ GeV and $v_s$ is real.

The scalar potential in Eq. (\ref{Z2 scalar potential}) is invariant under two different   $\mathbb{Z}_2$ symmetries,
\be
S\to -S,
\ee
and the CP-like
\be 
S\to S^\ast.
\ee  
After the symmetry breaking the latter symmetry becomes equivalent to $\chi\to -\chi$, which stabilizes $\chi$, making it a dark matter candidate. 

We will now move on to study the vacuum structure of the potential in Eq. (\ref{Z2 scalar potential}).

\subsection{Different vacua}
The potential in \eqref{Z2 scalar potential} is bounded from below if
\be 
\lambda_h > 0,\quad \lambda_s > 0, \quad 2\sqrt{\lambda_h \lambda_s}+\lambda_{hs} > 0, 
\ee
 There are three possible vacuum configurations in field space $(h,s,\chi)$: the mixed vacuum $(v_h, v_s,0)$, the Higgs vacuum $(v'_h, 0, 0)$ and the singlet vacuum $(0, v'_s, 0)$. Let us study if there can be degenerate minima.

\subsubsection{Our vacuum $(v_h,v_s,0)$}

In order for the pseudo-Goldstone to be the DM candidate, our Universe has to reside in the mixed vacuum. This vacuum $(v_h,v_s,0)$ satisfies the following minimization conditions:
\begin{align}
\mu_h^2 
& = -\lambda_h v_h^2-\frac{1}{2}\lambda_{hs} v_s^2\label{Our vacuum minimization condition h},\\
\mu_s^2 & = -\lambda_s v_s^2-\frac{1}{2}\lambda_{hs} v_h^2+\frac{1}{2}\mu^{\prime 2},
\label{Our vacuum minimization condition s}
\end{align}
which can be inverted in favor of the VEVs:
\begin{align}
v_h^2 & =  \frac{-4\lambda_s \mu_h^2 + 2\lambda_{hs}\mu_s^2 -\lambda_{hs}\mu^{\prime 2}}{4\lambda_h\lambda_s-\lambda_{hs}^2}>0\label{solved VEVs 1},\\
v_s^2 & = \frac{-4\lambda_h \mu_s^2 + 2\lambda_{hs}\mu_h^2 -2\lambda_{h}\mu^{\prime 2}}{4\lambda_h\lambda_s-\lambda_{hs}^2}>0.\label{solved VEVs 2}
\end{align}

The mass matrix in the $(h,s,\chi)$-basis is given by
\be
\mathcal{M}^2=\begin{pmatrix}
2\lambda_h v_h^2 & \lambda_{hs} v_h v_s & 0\\
\lambda_{hs} v_h v_s & 2\lambda_s v_s^2 & 0\\
0 & 0 & \mu^{\prime 2}
\end{pmatrix}.
\ee

The determinant of the $2\times 2$ CP-even block matrix has to be strictly positive in order for the mixed configuration to be a physical minimum, yielding the condition
\be\label{positive CP-even masses}
4\lambda_h\lambda_s-\lambda_{hs}^2 >0.
\ee
We have chosen    the parameter $\mu^{\prime 2}$  to be positive and therefore all the masses will be positive. The mixed vacuum configuration $(v_h, v_s,0)$ is therefore the minimum to which the other configurations are compared.

When one combines Eq. \eqref{positive CP-even masses} with Eqs. \eqref{solved VEVs 1} and \eqref{solved VEVs 2}, the following inequalities are obtained:
\begin{align}
-4\lambda_s\mu_h^2+2\lambda_{hs}\mu_s^2-\lambda_{hs}\mu^{\prime 2} &>0,\label{VEV positivity condition 1}\\
-4\lambda_h\mu_s^2+2\lambda_{hs}\mu_h^2+2\lambda_h\mu^{\prime 2} &>0.\label{VEV positivity condition 2}
\end{align}

\subsubsection{The Higgs vacuum $(v'_h,0,0)$}

While the Higgs vacuum $(v'_h,0,0)$ gives rise to correct electroweak symmetry breaking, in this case it is not the pseudo-Goldstone boson but the complex singlet $S$ that is the DM candidate, contrary to the present assumptions. The mass matrix in this vacuum is given by
\be
\mathcal{M}^2=\begin{pmatrix}
-2\mu_h^2 & 0 & 0\\
0 & -\frac{\lambda_{hs}\mu_h^2}{2\lambda_h}+\mu_s^2-\frac{\mu^{\prime 2}}{2} & 0\\
0 & 0 & -\frac{\lambda_{hs}\mu_h^2}{2\lambda_h}+\mu_s^2+\frac{\mu^{\prime 2}}{2}
\end{pmatrix}.
\ee
By using Eq. (\ref{VEV positivity condition 2}), one sees that the mass in the middle is negative. Therefore this vacuum configuration is not a minimum.

\subsubsection{The singlet vacuum $(0,v'_s,0)$}

The singlet vacuum $(0,v'_s,0)$ can in principle be a minimum. Let us compare, however the potential energy of this minimum to that of our mixed minimum $(v_h,v_s,0)$. Their difference is
\be
\begin{split}
  V(v_h,v_s,0)&-V(0,v'_s,0)  =-\frac{1}{4}\frac{1}{4\lambda_h\lambda_s-\lambda_{hs}^2} \frac{1}{\lambda_s}
  \\
  &\times \left[\frac{\lambda_{hs}}{2}(2\mu_s^2-\mu^{\prime 2})-2\lambda_s\mu_s^2\right]^2<0,
\end{split}
\ee
according to Eq.~(\ref{positive CP-even masses}). Therefore, we have that
\be
V(v_h,v_s,0)<V(0,v'_s,0).
\ee
This inequality is strict, and therefore the singlet minimum $(0,v'_s,0)$ cannot be degenerate with our  vacuum.

\subsection{Verdict on $\mathbb{Z}_2$ pGDM from the PMPC}
We see that if we demand  that the mixed vacuum $(v_h,v_s,0)$ be a minimum of the scalar potential, the other vacuum configurations cannot be true minima or if they are, they cannot be degenerate with our minimum. This is in a agreement with \cite{Azevedo:2018oxv}. Therefore, the singlet vacuum $(0,v'_s,0)$ cannot be a  minimum according to the PMPC. The PMPC allows for the $\mathbb{Z}_2$ pseudo-Goldstone dark matter model only if $(0, v'_s, 0)$ is not a minimum.

\section{Model III:  $\mathbb{Z}_3$ pseudo-Goldstone DM}
\label{sec:model3}

The $\mathbb{Z}_3$ pseudo-Goldstone DM model is an extension of SM with a complex scalar $S$ carrying a global $U(1)$ charge. The most general  scalar potential invariant under a global $U(1)$ symmetry is
\be\label{lagrangian}
\begin{split}
V_0 &= \mu_h^2 H^\dagger H+ \mu_s^2 S^\ast S+\lambda_h (H^\dagger H)^2+\lambda_s(S^\ast S)^2
\\ 
&+\lambda_{hs} (H^\dagger H)(S^\ast S),
\end{split}
\ee
where $H$ is the SM Higgs doublet. 
The global $U(1)$ symmetry is spontaneously broken as the scalar $S$ acquires a non-zero vacuum expectation value. This would result in a massless Goldstone boson in the physical spectrum. In order to avoid this, the global $U(1)$ symmetry is explicitly broken into its discrete subgroup $\mathbb{Z}_3$ by the following soft term:
\be
V_{\mathbb{Z}_3}=\frac{{\mu'_s}}{2}S^3+\text{h.c.},
\ee
so the full potential becomes
\be\label{Z3 scalar potential}
V=V_0+V_{\mathbb{Z}_3}.
\ee
The parameter ${\mu'_s}$ can be made real by absorbing its phase to the field $S$. When ${\mu'_s}$ is negative the VEV of $S$ is real and positive.  This can  be easily seen by writing the singlet in polar coordinates:
\be
S=\frac{1}{\sqrt{2}}r e^{i\theta}.
\ee
The soft term now becomes
\be
V_{\mathbb{Z}_3}=\frac{{\mu'_s}}{2}\frac{1}{2^{3/2}}r^3\cos (3\theta).
\ee
The soft term is the only part of the potential that depends on the phase. Therefore the global minimum of the potential is the minimum of the soft term. Let us solve the phase at the minimum. Since  ${\mu'_s}<0$, the minimum is obtained when $\cos(3\theta)=1$, which implies that
\be
3\theta =0+2\pi n\Rightarrow \theta=0,\frac{2\pi}{3},-\frac{2\pi}{3}.
\ee
Therefore, the potential has a three-fold degenerate minimum. One of them is real and two are complex. The different vacua are related to each other by $\mathbb{Z}_3$ transformations $S\rightarrow e^{2i\pi/3} S$. The singlet VEV can always be chosen to be real. Note that with the choice, ${\mu'_s}>0$, the VEV of $S$ is negative.

We now parametrize the scalar fields in unitary gauge as
\be
H=\frac{1}{\sqrt{2}}\begin{pmatrix}
0\\
h+v_h
\end{pmatrix},
\ee
and
\be
S=\frac{1}{\sqrt{2}}(v_s+s+i\chi).
\ee
The Higgs vacuum expectation value is $v_h=246$ GeV.
The pseudo-scalar $\chi$ can be dark matter \cite{Kannike:2019mzk}. The $\mathbb{Z}_3$-symmetric complex singlet model has also been studied in the phase where  the $\mathbb{Z}_3$ remains unbroken \cite{Belanger:2012zr,Hektor:2019ote}.  

The scalar potential in Eq. (\ref{Z3 scalar potential}) is invariant under the CP-like $\mathbb{Z}_2$-symmetry
\be
S\to S^\ast.
\ee
This symmetry is becomes 
\be 
\chi\to -\chi,
\ee
when $S$ acquires VEV. This symmetry stabilizes $\chi$, making it a dark matter candidate. 

We next  study the vacuum structure of the model.
 
\subsection{Different vacua}

The potential in Eq. (\ref{Z3 scalar potential}) is bounded from below, if
\be 
\lambda_h > 0,\quad \lambda_s > 0, \quad 2\sqrt{\lambda_h \lambda_s}+\lambda_{hs} > 0. 
\ee
  
There are three possible vacuum configurations in the field space $(h,s,\chi)$: the mixed vacuum $(v_h,v_s,0)$, the Higgs vacuum $(v'_h,0,0)$ and the singlet vacuum $(0,v'_s,0)$.

\subsection{Our vacuum $(v_h,v_s,0)$}

In order for the pseudo-Goldstone to be the DM candidate, our Universe has to reside in the mixed vacuum. 
The vacuum $(v_h,v_s,0)$ satisfies the following minimization conditions:
\bea
\mu_h^2 
& = & -\lambda_h v_h^2-\frac{1}{2}\lambda_{hs} v_s^2,
\label{Our vacuum minimization condition h}\\
\mu_s^2 & = &-\lambda_s v_s^2-\frac{1}{2}\lambda_{hs} v_h^2-\frac{3}{2\sqrt{2}}v_s{\mu'_s}.\label{Our vacuum minimization condition s}
\eea 
The minimization conditions are now hard to invert in favor of VEVs as minimization condition of $\mu_s^2$ contains a linear term in $v_s$, due to the soft cubic self-coupling of $S$.  The VEVs are solved to be:
\begin{align}
v_s & =  \frac{-\frac{3\mu'_s}{\sqrt{2}}\pm 2\sqrt{\frac{1}{\lambda_h}\left(4\lambda_h\lambda_s-\lambda_{hs}^2\right)\left(\frac{\lambda_{hs}\mu_h^2}{2\lambda_h}-\mu_s^2\right)+\frac{9\mu_{s}^{\prime 2}}{8}}}{\frac{1}{\lambda_h}\left(4\lambda_h\lambda_s-\lambda_{hs}^2\right)},
\label{vs our vacuum minimum}\\
v_h & =  \pm\frac{\sqrt{-\lambda_{hs} v_s^2-2\mu_h^2}}{\sqrt{2}\sqrt{\lambda_h}}.
\end{align}
We choose the positive sign $v_s$-solution to be the VEV in our minimum. The negative sign $v_s$-solution is not a minimum as one of the scalar masses is necessarily negative in this case.

The mass matrix in the $(h,s,\chi)$-basis is given by
\be\label{Z3 our minimum masses}
\mathcal{M}^2=\left(\begin{array}{ccc}
2\lambda_h v_h^2 & \lambda_{hs} v_h v_s & 0\\
\lambda_{hs} v_h v_s & 2\lambda_s v_s^2+\frac{3}{2\sqrt{2}}v_{s} \mu'_s & 0\\
0 & 0 & -\frac{9v_s \mu'_s}{2\sqrt{2}}
\end{array}
\right).
\ee

The $2 \times 2$ block in the upper-left corner of Eq. \eqref{Z3 our minimum masses} can be diagonalized by an orthogonal transformation
\be
U^T M^2 U = \textrm{diag}(m_h^2,m_H^2),
\ee
where
\be
M^2=\left(\begin{array}{cc}
2\lambda_h v_h^2 & \lambda_{hs} v_h v_s \\
\lambda_{hs} v_h v_s & 2\lambda_s v_s^2+\frac{3}{2\sqrt{2}}vs \mu'_s 
\end{array}
\right),
\ee
and
\be
U=\left(\begin{array}{cc}
\cos\theta & \sin\theta \\
-\sin\theta & \cos\theta 
\end{array}
\right).
\ee
The mixing angle $\theta$ is defined as
\be
\tan 2\theta =-\frac{\lambda_{hs}v_h v_s}{\lambda_h v_h^2-\lambda_s v_s^2-\frac{3}{4\sqrt{2}}v_s \mu'_s}.
\ee
With the knowledge that we must choose the positive sign solution of $v_s$, the parameters in the potential can be interchanged to more suitable ones. The potential parameters $\mu_h^2$, $\mu_s^2$, $\lambda_h$, $\lambda_s$, $\lambda_{hs}$ and $\mu'_s$ are replaced by $v_h$, $v_s$, $m_h$, $m_H$, $m_\chi$ and the mixing angle $\theta$:
\begin{align}
\mu_h^2 & = -\frac{1}{4}(m_h^2+m_H^2)+\frac{1}{4v_h}(m_H^2-m_h^2)
 \notag
 \\
 &\times (v_h\cos 2\theta -v_s\sin 2\theta),\\
\mu_s^2 & = -\frac{1}{4}(m_h^2+m_H^2)+\frac{1}{4v_s}(m_h^2-m_H^2)
\notag
\\
&\times(v_s\cos 2\theta +v_h\sin 2\theta)+\frac{1}{6}m_\chi^2,\\
\lambda_h & = \frac{m_h^2+m_H^2+(m_h^2-m_H^2)\cos 2\theta}{4 v_h^2},\\
\lambda_s & = \frac{3(m_h^2+m_H^2)+3(m_H^2-m_h^2)\cos 2\theta+2m_\chi^2}{12 v_s^2},\\
\lambda_{hs} & = -\frac{(m_h^2-m_H^2)\sin 2\theta}{2 v_h v_s},\\
\mu'_s & = -\frac{2\sqrt{2}}{9}\frac{m_\chi^2}{v_s}.
\end{align}

\subsection{The Higgs vacuum $(v'_h,0,0)$}

While the Higgs vacuum $(v'_h,0,0)$ gives rise to correct electroweak symmetry breaking, in this case it is not the pseudo-Goldstone boson but the complex singlet $S$ that is the DM candidate, contrary to the present assumptions.
The mass matrix in this vacuum is given by
\be
\mathcal{M}^2=\begin{pmatrix}
-2\mu_h^2 & 0 & 0\\
0 & -\frac{\lambda_{hs}\mu_h^2}{2\lambda_h}+\mu_s^2 & 0\\
0 & 0 & -\frac{\lambda_{hs}\mu_h^2}{2\lambda_h}+\mu_s^2
\end{pmatrix}.
\ee
All the masses can be positive simultaneously. This minimum can be degenerate with our mixed vacuum, leading into the condition
\be
\mu^{\prime 2}=\frac{1}{\lambda_h}\left(4\lambda_h\lambda_s-\lambda_{hs}^2\right)\left(-\frac{\lambda_{hs}}{2\lambda_h}\mu_h^2+\mu^2_s\right),
\ee
or equivalently,
\be\label{higgs only vacuum degeneracy}
m_\chi^2 = \frac{9m_h^2 m_H^2}{m_h^2\cos^2\theta+m_H^2\sin^2\theta}.
\ee

\subsection{The singlet vacuum $(0,v'_s,0)$}
In the singlet vacuum $(0,v'_s,0)$, only the complex singlet $S$ gets a VEV and electroweak symmetry is not broken.
The minimization condition is given by
\be
\mu_s^2=-\lambda_s {v'_s}^2-\frac{3}{2\sqrt{2}}v'_s \mu'_s.
\ee
The VEV $v'_s$ can be found to be
\be
v'_s=\frac{-3\mu'_s\pm \sqrt{-32\lambda_s\mu_s^2+9\mu^{\prime 2}}}{4\sqrt{2}\lambda_s}.
\ee

The mass matrix in this vacuum is given by
\be
\mathcal{M}^2=\left(\begin{array}{ccc}
\mu_h^2-\frac{\lambda_{hs}}{2 \lambda_s}\left[\mu_s^2+\frac{3\mu'_s v'_s}{2\sqrt{2}}\right] & 0 & 0\\
0 & -2\mu_s^2-\frac{3\mu'_s v'_s}{2\sqrt{2}} & 0\\
0 & 0 & -\frac{9\mu'_s v'_s}{2\sqrt{2}}
\end{array}
\right).
\ee

This vacuum configuration does not seem to be able to be a minimum when the mixed vacuum $(v_h,v_s,0)$ and the Higgs vacuum $(v'_h,0,0)$ are degenerate. It is hard to show analytically, but all numerical evidence points to that.

\begin{figure}[tb]
\begin{center}
\includegraphics[width=0.5\textwidth]{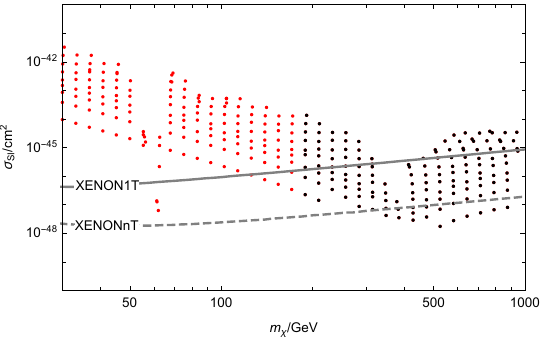}
\end{center}
\caption{Direct detection cross section for $\mathbb{Z}_{3}$ pGDM. The XENON1T result \cite{Aprile:2018dbl} is shown in solid line and prediction \cite{Ni2017} for the XENONnT in dashed line. The red points are excluded by the Higgs invisible width or a too large Higgs total width.}
\label{fig:Z3:direct:detection}
\end{figure}

\subsection{Final verdict on $\mathbb{Z}_3$ pGDM from the PMPC}
The two vacuum configurations $(v_h,v_s,0)$ and $(v'_h,0,0)$ can be minima and degenerate at the same time.  This degeneracy of these two minima will impose the condition in Eq. (\ref{higgs only vacuum degeneracy}), which aproximately reads:
\be
m_\chi\approx 3 m_H,
\ee
so strictly speaking, $\chi$ is on the borderline of not being a true pseudo-Goldstone boson.
The third vacuum configuration $(0,v'_s,0)$ does not seem to be a minimum when the $(v_h,v_s,0)$ and $(v'_h,0,0)$ are degenerate. This is also checked only numerically.

We also impose constraints from collider experiments. The fit of the Higgs signal strengths constrains the value of the mixing angle to $\cos^2\theta\ge 0.9$~\cite{Beacham:2019nyx}. If the mass of $x \equiv \chi \text{ or }h_2$ is less than $m_h/2$, then the Higgs decay width into this particle is given by \cite{Kannike:2019mzk}
\begin{equation}
  \Gamma_{h \to x x} = \frac{g_{hxx}}{8 \pi} \sqrt{1 - 4 \frac{m_x^2}{m_h^2}}
\end{equation}
with
\begin{align}
  g_{h1 \chi \chi} &= \frac{m_h^2 + m_\chi^2}{v_S},
  \\
  g_{h1 h_2 h_2} &= \frac{1}{v v_S} \left[ \left(\frac{1}{2} m_h^2 + m_2^2\right) (v \cos \theta + v_S \sin \theta) \right.
  \notag
  \\
  & \left. 
  + \frac{1}{6} v m_\chi^2 \cos \theta \right] \sin 2 \theta.
\end{align}
The latest measurements of the width of an SM-like Higgs boson
gives $\Gamma^{h_1}_\text{tot}\,=3.2^{+2.8}_{-2.2}$ MeV, with 95\% CL limit on
$\Gamma_\text{tot} \leq 9.16$ MeV \cite{Sirunyan:2019twz}.
In addition, the Higgs invisible branching ratio is then given by
\begin{equation}
\text{BR}_\text{inv} = \frac{\Gamma_{h \to \chi\chi}}{\Gamma_{h_1 \to \text{SM}} + \Gamma_{h \to \chi\chi} + \Gamma_{h \to h_2 h_2}},
\end{equation}
which is constrained to be below $0.24$ at $95\%$ confidence level \cite{Khachatryan:2016whc,ATLAS-CONF-2018-031} by direct measurements and below about $0.17$ by statistical fits of Higgs couplings \cite{Giardino:2013bma,Belanger:2013xza}. Decays of the resulting $h_{2}$ into the SM may be visible \cite{Huitu:2018gbc}.

A scan of the parameter space with the condition \eqref{higgs only vacuum degeneracy} imposed is shown in Fig.~\ref{fig:Z3:direct:detection}. We take the mass of the other CP-even scalar $m_{2} \in [10, 1500]$ and the mixing angle $\sin \theta \in \pm [0.01, 0.3]$. The VEV $v_{s}$ is fitted from the DM relic density $\Omega_{c} h^{2} = 0.120$ \cite{Aghanim:2018eyx}. The points in red are excluded by the Higgs invisible width and/or by an overly large Higgs total width.

\section{Discussion and Conclusion}
\label{sec:conc}

In this work we have argued that the principle of multiple point criticality must be generalized to all vacuum states, implying that all minima of a scalar potential 
must be degenerate. Indeed, if the PMPC follows from some fundamental underlying principle, its validity must not depend on the particulars of the model.
For example, in the most studied case, one minimum is taken to be the SM EW Higgs minimum and another one the Planck-scale minimum generated by the running of Higgs quartic coupling. We argue that the principle must apply to all minima in multi-scalar models, whether they are generated radiatively or at tree level.

Phenomenological implications of this formulation of the PMPC are straightforward but profound -- the PMPC becomes a tool to constrain the parameter space of 
multi-scalar models. Since the scalar singlet models represent one of the most popular class of dark matter models, we exemplified the usefulness of the PMPC by studying how different realizations of scalar DM models are constrained  by imposing the PMPC requirements. 

In the simplest possible model, studied in Section~\ref{sec:model1}, the DM is a real scalar singlet stabilized by a $\mathbb{Z}_2$ symmetry. The PMPC predicts that the singlet self-coupling hits a Landau pole at DM mass of about $300$~GeV, invalidating the model above that scale. Due to the stringent direct detection constraints, the DM mass in this model is predicted to be near the Higgs resonance.

The second model under investigation, the $\mathbb{Z}_2$ symmetric pseudo-Goldstone DM model, studied in Section~\ref{sec:model2}, cannot have degenerate minima at all. Therefore it cannot be  consistent with the PMPC requirements. 

The third model under investigation, the $\mathbb{Z}_3$ symmetric pseudo-Goldstone DM model, studied in Section~\ref{sec:model3}, can have degenerate minima. Applying the PMPC in this case  leads to a simple relation between the DM and the second scalar masses and excludes DM masses below about $190$~GeV via the measured Higgs width. This model represents an example how the PMPC can constrain the phenomenology of multi-scalar models.

In conclusion, if the PMPC is indeed realized in Nature, which we have assumed in this work, it may lead to significant constraints and simplifications of the phenomenology of multi-scalar models. Further studies of degenerate tree-level minima in those models is worth of pursuing.


\vspace{5mm}
\noindent \textbf{Acknowledgement.} We thank Gia Dvali, Luca Marzola, 
Alessandro Strumia, Daniele Teresi and Hardi Veerm{\"a}e for discussions on the origin of PMPC. 
This work was supported by the Estonian Research Council grants PRG434, PRG803, PRG803, MOBTT5 and MOBTT86, and by the EU through the European Regional Development Fund CoE program TK133 ``The Dark Side of the Universe." 


\bibliography{multiple_point_criticality_principle}

\end{document}